# On the relative role of different age groups during epidemics associated with the respiratory syncytial virus

E. Goldstein[1,*], H. H. Nguyen[1], P. Liu[2], C. Viboud[3,**], C.A. Steiner[4], C. J. Worby[5], M. Lipsitch[1,6]

1. Center for Communicable Disease Dynamics, Department of Epidemiology, Harvard TH Chan School of Public Health, Boston, MA 02115
2. Department of Applied Mathematics, Harvard University, Cambridge, MA 02138
3. Division of International Epidemiology and Population Studies, Fogarty International Center, National Institutes of Health, Bethesda, MD 20892
4. Agency for HealthCare Research and Quality, U.S. Department of Health & Human Services, Rockville, MD 20850 - Currently works at Institute for Health Research, Kaiser Permanente Colorado, Denver, CO 80231
5. Woodrow Wilson School of Public and International Affairs, Princeton University, Princeton, NJ 08544
6. Department of Immunology and Infectious Diseases, Harvard TH Chan School of Public Health, Boston, MA 02115

* Corresponding author; Email: egoldste@hsph.harvard.edu
** Second corresponding author; Email: viboudc@mail.nih.gov

## Abstract

*Background:* While RSV circulation results in high burden of hospitalization, particularly among infants, young children and the elderly, little is known about the role of different age groups in propagating annual RSV epidemics in the community.

*Methods:* During a communicable disease outbreak, some subpopulations may play a disproportionate role during the outbreak's ascent due to increased susceptibility and/or contact rates. Such subpopulations can be identified by considering the proportion that cases in a subpopulation represent among all cases in the population occurring before (Bp) and after the epidemic peak (Ap) to calculate the subpopulation's relative risk, RR=Bp/Ap. We estimated RR for several age groups using data on RSV hospitalizations in the US between 2001-2012 from the Healthcare Cost and Utilization Project (HCUP).

*Results:* Children aged 3-4y and 5-6y each had the highest RR estimate for 5/11 seasons in the data, with RSV hospitalization rates in infants being generally higher during seasons when children aged 5-6y had the highest RR estimates. Children aged 2y had the highest RR estimate during one season. RR estimates in infants and individuals aged 11y and older were mostly lower than in children aged 1-10y.

*Conclusions:* The RR estimates suggest that preschool and young school-age children have the leading relative roles during RSV epidemics. We hope that those results will aid in the design of RSV vaccination policies.

**Introduction**

Annual RSV epidemics take place in the US, exerting a heavy toll in severe outcomes in infants and young children [1-4], with rates of associated bronchiolitis hospitalizations in infants being particularly high [5,6]. Rates of hospitalizations with RSV present in the diagnosis drop off dramatically with age ([7], Table 1 – see also Table 1 in our paper), but those represent only a fraction of all hospitalizations associated with RSV infections [8], and our understanding of the hospitalization burden stemming from the circulation of the respiratory syncytial virus is limited, particularly for the adult populations [9,10]. Previous work [11] showed high rates of RSV-associated respiratory hospitalizations among older adults (see also [7] and [12]).

Given the high burden of severe outcomes associated with RSV, it would be useful to understand the roles of different population groups in propagating RSV epidemics in the community. Such information should contribute to our overall understanding of RSV epidemiology that should help inform mitigation efforts, with a variety of RSV vaccine candidates for different populations being in different stages of development [13,14]).

In [15-17] we developed a method for assessing the roles of different subpopulations in transmission during epidemics of infectious diseases. The idea of that method is that subpopulations that play a disproportionate role during the outbreak's ascent due to increased susceptibility and/or contact rates can be related to the relative risk (RR) statistic that evaluates the change in the subpopulation's proportion among all cases in the population before vs. after the epidemic's peak (see Methods). In [15], we have estimated the RR statistic in select age groups for several influenza epidemics in the US using data on hospitalized cases for the inference. An important aspect of the method proposed in [15] is that the results do not depend on case-hospitalization rates in different age groups. Moreover, we have used simulations to show a relation between a higher value for an RR statistic in a given age group and a higher impact of vaccinating an individual in that age group on reducing the epidemic's initial growth rate.

In this paper we estimate the RR statistic in different age groups for eleven RSV epidemics in the US using data on RSV hospitalizations in a collection of US states between 2001 and 2012 [18]. We consider a fine age stratification for young children to examine how their role during RSV epidemics changes with age progression. We compare the RR estimates in different age groups across eleven RSV epidemics and discuss the relevance of our findings to the study of a potential impact of vaccination.

**Methods**

*Data*
We used weekly data on hospitalizations with an RSV diagnosis (both primary and contributing [*ICD-9*] codes 079.6, 466.11, 480.1) from the State Inpatient Databases of the Healthcare Cost and Utilization Project (HCUP), maintained by the Agency for Healthcare Research and Quality through an active collaboration [18]. This database contains hospital discharges from community hospitals in participating states. We compiled data for the 2001-2002 through the 2011-2012 RSV seasons for 10 different age groups and 26 participating states (representing about 54.3% of the US population between 2001-2012) that reported continuously to HCUP between week 27, 2001 and week 26, 2012. Those states are: AZ, CA, CO, CT, GA, IA, IL, KY, MD, MN, MO, NC, NE, NJ, NV, OH, OR, RI, SC, SD, TN, UT, VT, WA, WI, WV. Each RSV season was defined as a period between calendar week 27 of one year through calendar week 26 on the next year. One year (2006) had 53 calendar weeks; the corresponding season was defined as the period from week 27 of 2006 through week 25 of 2007.

*Before-and-after the peak counts*
Data on the cumulative number/rate per 100,000 of RSV hospitalizations in different age groups during each season (Table 1) suggest the following age splitting (in years) for the analysis: (<1,1,2,3-4,5-6,7-10,11-17,18-49,50-64,65+). For each season and each state $s$, we defined the peak week $P(s)$ for that state as the calendar week with the highest total number of RSV hospitalizations in the state. Each case in the state during the given season was classified as before-the-peak case if it occurred in or prior to week $P(s) - 2$; after-the-peak case if it occurred in or after week $P(s) + 2$. We exclude cases occurring between weeks $P(s) - 1$ and $P(s) + 1$ from the analysis to avoid misclassification of counts as before or after-the-peak stemming from the noise in the count data for the hospitalized cases, differences in *case-hospitalization rates* (proportion of RSV infections that result in hospitalizations with an RSV diagnosis) for the various age groups that may result in different peak times for the counts of the hospitalized cases vs. cases of infection in the community, etc.

For each age group $g$, before-and-after the peak seasonal counts in that age group in different states were combined into national before-and-after the peak counts $B(g)$ and $A(g)$ correspondingly.

*Statistical Inference*
The point estimate for the seasonal relative risk $RR(g)$ in an age group $g$ is

$$\frac{B(g)}{\sum_h B(h)} \bigg/ \frac{A(g)}{\sum_h A(h)}$$

The estimates and confidence bounds for relative risks in each group were obtained in a Bayesian framework following the methodology in [10]. Briefly, as the number of cases of infection in each age group is large and case-hospitalization rates (see the previous sub-section) are low, we assume that the numbers of reported cases in each age group before an after the peak are Poisson distributed. Posterior samples for the Poisson parameters (with a flat prior) corresponding to the counts $B(h), A(h)$ are generated; for each i=1,..,100000, the corresponding parameters $B^i(h), A^i(h)$ are plugged into the equation above to generate an estimate $RR^i(g)$ for the relative risk. The mean and the credible interval for the sample $(RR^i(g))$ (i=1,..,100000) are then extracted.

**Results**

Table 1 presents the cumulative number and the rate per 100,000 of RSV hospitalizations during each season in each age group for the 26 states that reported continuously between the 2001-2002 through 2011-2012 seasons (Methods). There is an apparent upward trend in the counts/rates in all age groups over 3 years of age (being most pronounced in adults), likely indicative of temporal changes in detection/diagnostic practices. We note that differences in hospitalization counts/rates by age need not correspond to differences in RSV infection rates by age – see also the 1st paragraph of the Discussion.

| Age / Season | 2001-02 | 2002-03 | 2003-04 | 2004-05 | 2005-06 | 2006-07 | 2007-08 | 2008-09 | 2009-10 | 2010-11 | 2011-12 |
|---|---|---|---|---|---|---|---|---|---|---|---|
| <1y | 44855/2049 | 38518/1766 | 40855/1856 | 34133/1546 | 38272/1725 | 38346/1697 | 39289/1721 | 34462/1542 | 37810/1734 | 37126/1715 | 28692/1334 |
| 1y | 8677/400.9 | 7245/331.4 | 8213/377.9 | 7040/321.5 | 7436/338.5 | 7372/334.4 | 7986/356.2 | 7328/324.2 | 8921/402.9 | 8775/403.3 | 6757/311 |
| 2y | 2825/134 | 2267/104.8 | 2713/124.2 | 2343/108 | 2637/120.4 | 2485/113.1 | 2869/130.2 | 2690/120.1 | 3601/159.3 | 3645/164.3 | 2801/128.4 |
| 3y | 1059/50.6 | 908/43 | 1128/52.1 | 930/42.6 | 1069/49.1 | 1038/47.3 | 1112/50.4 | 1220/55.1 | 1727/76.9 | 1764/78 | 1364/61.4 |
| 4y | 437/20.8 | 351/16.7 | 526/24.8 | 476/21.9 | 465/21.1 | 419/19.1 | 490/22.1 | 489/22 | 790/35.3 | 829/36.7 | 661/29.2 |
| 5y | 209/9.8 | 161/7.6 | 226/10.7 | 207/9.7 | 233/10.7 | 197/8.9 | 237/10.7 | 274/12.3 | 399/17.8 | 452/20.1 | 320/14.1 |
| 6y | 118/5.4 | 88/4.1 | 116/5.5 | 102/4.9 | 117/5.5 | 109/5 | 133/6 | 146/6.6 | 209/9.4 | 257/11.5 | 200/8.9 |
| 7y | 66/3 | 56/2.6 | 58/2.7 | 66/3.1 | 61/2.9 | 57/2.7 | 67/3.1 | 96/4.3 | 124/5.6 | 160/7.2 | 88/3.9 |
| 8y | 48/2.2 | 44/2 | 36/1.6 | 34/1.6 | 42/2 | 45/2.1 | 45/2.1 | 51/2.3 | 73/3.3 | 88/4 | 76/3.4 |
| 9y | 40/1.8 | 24/1.1 | 20/0.9 | 29/1.3 | 36/1.7 | 36/1.7 | 44/2.1 | 49/2.3 | 85/3.8 | 55/2.5 | 34/1.5 |
| 10y | 31/1.3 | 28/1.2 | 36/1.6 | 24/1.1 | 35/1.5 | 32/1.4 | 37/1.6 | 43/1.9 | 55/2.4 | 57/2.5 | 48/2.1 |
| 11y | 33/1.4 | 27/1.2 | 19/0.8 | 25/1.1 | 31/1.4 | 19/0.8 | 19/0.8 | 38/1.7 | 48/2.1 | 60/2.6 | 40/1.7 |
| 12-17y | 106/0.8 | 84/0.6 | 106/0.8 | 81/0.6 | 109/0.8 | 116/0.8 | 142/1 | 151/1.1 | 194/1.4 | 218/1.6 | 149/1.1 |
| 18-49y | 191/0.3 | 132/0.2 | 177/0.2 | 150/0.2 | 238/0.3 | 235/0.3 | 318/0.4 | 373/0.5 | 493/0.7 | 568/0.8 | 465/0.6 |
| 50-64y | 156/0.7 | 120/0.5 | 144/0.6 | 162/0.6 | 267/1 | 219/0.8 | 320/1.1 | 328/1.1 | 616/2 | 723/2.2 | 564/1.7 |
| 65+y | 297/1.6 | 190/1 | 252/1.3 | 190/1 | 406/2.1 | 327/1.7 | 573/2.8 | 663/3.2 | 950/4.5 | 1331/6.2 | 1174/5.3 |

**Table 1:** Seasonal counts and rates per 100,000 for hospitalizations with RSV mentioned in the diagnosis in select age groups (in years). Data from the State Inpatient Databases of the Healthcare Cost and Utilization Project [18].

Tables 2 and 3 present the seasonal estimates of the relative risks in select age groups. Children aged 3-4y and 5-6y each had the highest RR estimate for 5/11 seasons in the data, with RSV hospitalization rates in infants being generally higher during seasons when children aged 5-6y had the highest RR estimates (Tables 1,2). Children aged 2y had the highest RR estimate during one season, with RR estimates in children aged 2y being higher than the ones in children aged 1y in 9/11 seasons. RR estimates in infants and individuals aged 11y and older were mostly lower than in children aged 1-10y.

| Season / Age | <1 | 1y | 2y | 3-4y | 5-6y |
|---|---|---|---|---|---|
| 2001-02 | 0.964 (0.95,0.97) | 1.172 (1.12,1.23) | 1.181 (1.08,1.28) | 1.146 (1.01,1.29) | 1.284 (0.99,1.64) |
| 2002-03 | 0.94 (0.93,0.95) | 1.211 (1.15,1.27) | 1.33 (1.21,1.46) | 1.312 (1.16,1.48) | 1.41 (1.06,1.85) |
| 2003-04 | 0.93 (0.92,0.94) | 1.206 (1.15,1.26) | 1.272 (1.17,1.39) | 1.396 (1.24,1.57) | 2.103 (1.6,2.73) |
| 2004-05 | 0.945 (0.93,0.96) | 1.121 (1.07,1.18) | 1.369 (1.25,1.5) | 1.408 (1.25,1.58) | 1.363 (1.04,1.76) |
| 2005-06 | 0.935 (0.92,0.95) | 1.194 (1.14,1.25) | 1.476 (1.35,1.62) | 1.534 (1.36,1.73) | 1.386 (1.08,1.76) |
| 2006-07 | 0.949 (0.94,0.96) | 1.164 (1.11,1.22) | 1.29 (1.18,1.41) | 1.473 (1.31,1.65) | 1.398 (1.09,1.77) |
| 2007-08 | 0.963 (0.95,0.97) | 1.184 (1.13,1.24) | 1.296 (1.19,1.41) | 1.217 (1.08,1.36) | 1.176 (0.93,1.47) |
| 2008-09 | 0.989 (0.98,1) | 1.113 (1.06,1.17) | 1.104 (1.01,1.2) | 1.19 (1.07,1.32) | 0.978 (0.78,1.21) |
| 2009-10 | 0.905 (0.89,0.92) | 1.271 (1.21,1.33) | 1.259 (1.17,1.35) | 1.504 (1.37,1.65) | 1.721 (1.42,2.08) |
| 2010-11 | 0.973 (0.96,0.99) | 1.164 (1.11,1.22) | 1.215 (1.13,1.31) | 1.35 (1.23,1.48) | 1.391 (1.16,1.65) |
| 2011-12 | 0.942 (0.93,0.96) | 1.178 (1.12,1.24) | 1.297 (1.19,1.41) | 1.425 (1.29,1.57) | 1.22 (1,1.48) |

**Table 2:** Seasonal RR estimates for select age groups (children under 7y) for RSV epidemics between 2001-2012 (Methods). Data from the State Inpatient Databases of the Healthcare Cost and Utilization Project [18].

| Season / Age | 7-10y | 11-17y | 18-49y | 50-64y | 65+y |
|---|---|---|---|---|---|
| 2001-02 | 0.733 (0.51,1.01) | 0.738 (0.5,1.05) | 0.65 (0.46,0.89) | 0.926 (0.62,1.33) | 0.382 (0.28,0.51) |
| 2002-03 | 0.953 (0.66,1.33) | 1.089 (0.7,1.62) | 0.973 (0.66,1.39) | 0.929 (0.62,1.34) | 0.812 (0.57,1.13) |
| 2003-04 | 1.732 (1.17,2.49) | 1.599 (1.02,2.42) | 0.816 (0.57,1.12) | 0.993 (0.68,1.41) | 1.053 (0.8,1.36) |
| 2004-05 | 1.34 (0.92,1.89) | 1.014 (0.65,1.5) | 0.739 (0.5,1.05) | 0.748 (0.51,1.06) | 0.605 (0.42,0.84) |
| 2005-06 | 1.178 (0.82,1.65) | 0.869 (0.57,1.27) | 0.711 (0.53,0.94) | 0.606 (0.46,0.79) | 0.584 (0.46,0.73) |
| 2006-07 | 1.294 (0.92,1.78) | 0.757 (0.49,1.1) | 0.693 (0.51,0.92) | 0.48 (0.34,0.65) | 0.493 (0.37,0.64) |
| 2007-08 | 0.946 (0.67,1.29) | 0.718 (0.49,1.01) | 0.516 (0.39,0.67) | 0.57 (0.43,0.74) | 0.389 (0.31,0.48) |
| 2008-09 | 1.139 (0.84,1.51) | 0.612 (0.43,0.84) | 0.566 (0.44,0.71) | 0.566 (0.43,0.72) | 0.511 (0.42,0.61) |
| 2009-10 | 1.636 (1.26,2.09) | 1.124 (0.82,1.5) | 0.822 (0.67,1) | 0.747 (0.62,0.89) | 0.61 (0.52,0.71) |
| 2010-11 | 1.031 (0.8,1.3) | 0.749 (0.56,0.98) | 0.578 (0.47,0.7) | 0.593 (0.5,0.7) | 0.413 (0.36,0.47) |
| 2011-12 | 1.359 (1.02,1.79) | 0.884 (0.63,1.21) | 0.847 (0.69,1.03) | 0.591 (0.49,0.71) | 0.596 (0.52,0.68) |

**Table 3:** Seasonal RR estimates for select age groups (individuals over the age of 7y) for RSV epidemics between 2001-2012 (Methods). Data from the State Inpatient Databases of the Healthcare Cost and Utilization Project [18].

## Discussion

Annual RSV epidemics in the US exert a heavy burden of severe disease in young children, particularly infants [1-4]. High rates of RSV-associated hospitalization were also estimated for the elderly [7,11,12]. At the same time, as disease severity is highly age-specific, there is limited information about the differences in RSV infection rates by age, and the role that the different age groups play in the dynamics of RSV infections in the community. In this paper we examine the role of different age groups in propagating annual RSV epidemics in the US using the RR statistic introduced in our earlier work [15-17], with values for that statistic in various age groups related to the impact of vaccination of individuals in those age groups on the epidemic's initial growth rate. Our results suggest that the highest estimates for the RR statistic belonged to either children aged 3-4y or 5-6y for most RSV epidemics. Moreover, rates of hospitalizations in infants with RSV present in the diagnosis were generally higher during those seasons when children aged 5-6y had the highest estimate for the RR statistic. The estimates for the RR statistic for infants and individuals over the age of 11y were generally lower than the ones for children aged 1-10y.

While several vaccine candidates for the infant and the pediatric populations are currently in different stages of development [13,14], the schedule for RSV vaccine administration in non-infant populations is unclear. Such schedule should account for both the direct protection imparted by vaccines, as well as the impact of vaccination on RSV transmission in the community. We hope that our study makes a contribution towards examining those issues, particularly in pointing to the prominence of children aged 3-6y during RSV epidemics.

The estimate of the relative role of infants under one year of age may be skewed by several factors. First, the immunological status of some infants changes through the course of RSV epidemics due to the waning immunity rendered by maternal antibodies [19-21]. Secondly, their population changes during the course of the outbreak through new births and ageing, namely infants who are close to the age of one at the start of the outbreak would be over the age of one during the outbreak's descent. Both phenomena would distribute infections among infants more uniformly during the epidemic periods and bias the estimates of the RR statistic toward the null value of 1. While we cannot estimate the magnitude of that bias from the available data, we note that the value of the RR statistic for infants is consistently under 1 (Table 2), with values over 1 for the leading driver groups. This combination of RR estimates below 1 and biases towards the null value of 1 for infants makes it seem unlikely that in the absence of biases, the values of the RR statistic in infants during different seasons would be as high as those for the leading driver groups. One possible explanation behind the relatively low RR estimates for infants is that infants, while potentially exceptionally susceptible to infection, have limited contacts, particularly with their peers, compared with other children, with outbreaks in daycare, school, and other setting where those children interact likely contributing to the prominence of those children during RSV epidemics as suggested by our estimates.

Our study has some additional limitations. The relation between the RR statistic and the role played by an average individual in a given age group in transmission dynamics is not entirely clear. Our earlier work ([15,16]) had attempted to address this issue through simulations of transmission dynamics, finding an association between the RR statistic in a given age group and the impact of vaccination of an individual in that age group on the epidemic's initial growth rate/reproductive number. While we have used state-specific peaks of RSV hospitalizations to characterize the before-vs.-after the peak cases, further asynchrony in RSV epidemics within each state may make this categorization of some cases for the corresponding local outbreaks inaccurate. We note that this would bias all the RR estimates toward the null value of 1, whereas the data allows for a delineation of the driver groups. For the 2003-04 and 2009-10 seasons (that had the two largest influenza epidemics during the study period), values of the RR statistic for RSV epidemics for school-age children were the highest during the study period. There are several possible explanations for this including immunological interference between influenza and RSV, and changes in diagnosis/laboratory testing practices through the course of a season, possibly stemming from changes in perception regarding the potential viral etiologies for the hospitalized cases. The latter might introduce a bias in the RR estimates for a number of RSV seasons, and the scope of that bias is difficult to assess with the available data. We also note that annual rates of RSV hospitalization in different age groups (Table 1) suggest temporal (year-to-year) changes in testing practices; however this should not bias the RR estimates that are derived separately during each season, unless testing/diagnostic practices change through the course of the season in a manner that is not uniform for all age groups [15,17].

We believe that despite the above limitations, our study sheds new light on the role of different age groups during RSV epidemics, a subject that had received limited attention in the literature. It suggests the prominence of children aged 3-6y in propagating RSV epidemics, pointing to the potential benefit of RSV vaccine administration in that age group for mitigating RSV epidemics in the community.

**Acknowledgement**: We would like to thank the HCUP Partner states that voluntarily provide their data to the project, and without whom this research would not be possible (https://www.hcup-us.ahrq.gov/partners.jsp). We are also grateful to Zeynal Karaca, PhD, and Jenny Schnaier who helped us with this project on behalf of AHRQ.